# Gamow-Teller strength distributions and electron capture rates for $^{55}$Co and $^{56}$Ni.

**Jameel-Un Nabi[*], Muneeb-ur Rahman**
Faculty of Engineering Sciences, GIK Institute of Engineering Sciences and
Technology, Topi 23460, N.W.F.P., Pakistan

**Abstract.** The Gamow-Teller strength (GT) distributions and electron capture rates on $^{55}$Co and $^{56}$Ni have been calculated using the proton-neutron quasiparticle random phase approximation theory. We calculate these weak interaction mediated rates over a wide temperature ($0.01 \times 10^9 - 30 \times 10^9$ K) and density ($10 - 10^{11}$ g cm$^{-3}$) domain. Electron capture process is one of the essential ingredients involved in the complex dynamics of supernova explosion. Our calculations of electron capture rates show   differences with the reported shell model diagonalization approach calculations and are comparatively enhanced at presupernova temperatures. We note that the GT strength is fragmented over many final states.

Massive stars (M > 10M$_\odot$) end their life in gravitational collapse of their core and formation of a neutron star or a black hole by supernova explosion. The structure of the progenitor star, including that of its core, plays a substantial role in the development of the explosion process. Indeed, the efforts to simulate the explosion numerically are found to make a substantial difference in the ultimate outcome, depending upon the progenitor models. Because the final outcome of the explosion depends so sensitively on a variety of physical inputs at the beginning of each stage of the entire process (i.e., collapse, shock formation, and shock propagation), it is desirable to calculate the presupernova stellar structure with the best possible physical data and inputs currently available. The energy budget would be balanced in favor of an explosion by a smaller precollapse iron core mass.

The evolution of the massive stars and the concomitant nucleosynthesis has been the subject of much computation [1]. During the later part of their burning cycles, these stars develop an iron core and lack further nuclear fuels (any transformation of the strongly-bound iron nuclei is endothermic). The core steadily becomes unstable and implodes as result of free-electron captures and iron photodisintegration.

The collapse is very sensitive to the entropy and to the number of leptons per baryon, $Y_e$ [2]. These two quantities are mainly determined by weak interaction processes, namely electron capture and $\beta$ decay. The simulation of the core collapse is very much dependent on the electron capture of heavy nuclides [3].  In the early stage of the collapse $Y_e$ is reduced as electrons are captured by Fe peak nuclei.  The late evolution stages of massive stars are strongly influenced by weak interactions which act to determine the core entropy and electron to baryon ratio, $Y_e$, of the presupernova star, and hence its Chandrasekhar mass which is proportional to $Y_e^2$[4]. Electron capture reduces the number of electrons available for pressure support, while beta decay acts in the opposite direction. Both processes produce neutrinos which, for densities $\rho \leq 10^{11}$ g cm$^{-3}$, escape the star carrying away energy and entropy from the core. Electron capture and beta decay during the final evolution of a massive star are dominated by Fermi and Gamow-Teller (GT) transitions. In the astrophysical scenario nuclei are fully ionized so one has continuum electron capture from the degenerate electron plasma. The energies of the electrons are high enough to induce transitions to the GT resonance.

\* Corresponding author
e-mail: Jameel@giki.edu.pk
Phone: 0092-938-71858(ext. 2535), Fax: 0092-938-71862

Electron capture rates are very sensitive to the distribution of the GT$_+$ strength (in the GT$_+$ strength, a proton is changed into a neutron). GT$_+$ strength distributions on nuclei in the mass range A = 50-65 have been studied experimentally via (n, p) charge-exchange reactions at forward angles. Some were also being measured [e.g. 5-9]. Results show that, in contrast to the independent particle model, the total GT$_+$ strength is quenched and fragmented over many final states in the daughter nucleus caused by the residual nucleon-nucleon correlations. Both these effects are caused by the residual interaction among the valence nucleons and an accurate description of these correlations is essential for a reliable evaluation of the stellar weak interaction rates due to the strong phase space energy dependence, particularly of the stellar electron capture rates.

Recognizing the vital role played by the electron capture process, Fuller *et .al* (referred as FFN) [10] estimated systematically the rates for nuclei in the mass range A= 45-60 stressing on the importance of capture process to the GT giant resonance. The basic calculation was performed using a zero-order shell model code. The calculations of FFN have shown that for the densities above $10^7$ g cm$^{-3}$, electron capture transitions to the GT resonance are an important part of the rate.

The FFN rates were then updated taking into account quenching of GT strength by an overall factor of two by Aufderheide and collaborators [11]. They also compiled a list of important nuclides which affect $Y_e$ via the electron capture processes. They ranked $^{55}$Co and $^{56}$Ni the most important nuclei with respect to their importance for the electron capture process for the early presupernova collapse.

We account here the microscopic calculation of electron capture rates in the stellar matter for the nuclei $^{55}$Co and $^{56}$Ni using the proton-neutron quasiparticle random phase approximation (pn-QRPA) theory.

The pn-QRPA theory [12-14] has been shown to be a good microscopic theory for the calculation of beta decay half lives far from stability [14, 15]. The pn-QRPA theory was also successfully employed in the calculation of $\beta^+$/electron capture half lives and again satisfactory comparison with the experimental half-lives were reported [16]. The pn-QRPA theory was then extended to treat transitions from nuclear excited states [17]. In view of success of the pn-QRPA theory in calculating terrestrial decay rates, Nabi and Klapdor used this theory to calculate weak interaction mediated rates and energy losses in stellar environment for *sd*- [18] and *fp/fpg*-shell nuclides [19]. Reliability of the calculated rates was also discussed in detail in [19]. There the authors compared the measured data of thousands of nuclides with the pn-QRPA calculations and got good comparison (See also [20]). Here we use this extended model to calculate the electron capture rates in stellar matter for $^{55}$Co and $^{56}$Ni pertaining to presupernova and supernova conditions. The main advantage of using the pn-QRPA theory is that we can handle large configuration spaces, by far larger than possible in any shell model calculations. We include in our calculations parent excitation energies well in excess of 10 MeV (compared to a few MeV tractable by shell model calculations). In our model, we considered a model space up to *7* major shells.

Our Hamiltonian, $H^{QRPA} = H^{sp} + V^{pair} + V^{ph}_{GT} + V^{pp}_{GT}$, is diagonalized in three consecutive steps. Single particle energies and wave functions are calculated in the Nilsson model [21], which takes into account nuclear deformations. Pairing is treated in the BCS approximation. The proton-neutron residual interactions occur in two different forms, namely as particle-hole and particle-particle interaction. The interactions are given separable form and are characterized by two interaction constants $\chi$ and $\kappa$, respectively. The selections of these two constants are done in an optimal fashion. Details of the model parameters can be seen in [16, 22]. In this work, we took $\chi = 0.2$ MeV and $\kappa = 0.007$ MeV for $^{55}$Co. The corresponding values for $^{56}$Ni were 0.5 MeV and 0.065 MeV, respectively. Q values were taken from [23].

The weak decay rate from the *i*th state of the parent to the *j*th state of the daughter nucleus is given by

$$\lambda_{ij} = \ln 2 \frac{f_{ij}(T, \rho, E_f)}{(ft)_{ij}} \ ,$$

where $(ft)_{ij}$ is related to the reduced transition probability $B_{ij}$ of the nuclear transition by $(ft)_{ij} = D / B_{ij}$. D is a constant and $B_{ij}$'s are the sum of reduced transition probabilities of the Fermi and GT transitions. The phase space integral $(f_{ij})$ is an integral over total energy and for electron capture it is given by

$$f_{ij} = \int_{w_l}^{\infty} w \sqrt{w^2 - 1} (w_m + w)^2 F(+Z, w) G_- dw.$$

In the above equation, $w$ is the total energy of the electron including its rest mass, and $w_l$ is the total capture threshold energy (rest + kinetic) for electron capture. $G_-$ ($G_+$) is the electron (positron) distribution function.

The number density of electrons associated with protons and nuclei is $\rho Y_e N_A$ ( $\rho$ is the baryon density, and $N_A$ is Avogadro's number).

$$\rho Y_e = \frac{1}{\pi^2 N_A} (\frac{m_e c}{\hbar})^3 \int_0^{\infty} (G_- - G_+) p^2 dp \ .$$

Here $p = (w^2 - 1)^{1/2}$ is the electron momentum and the equation has the units of *mol cm⁻³*. This equation is used for an iterative calculation of Fermi energies for selected values of $\rho Y_e$ and $T$. Details of the calculations can be found in [18]. We did incorporate experimental data wherever available to strengthen the reliability of our rates. The calculated excitation energies (along with their log*ft* values) were replaced with the measured one when they were within 0.5 MeV of each other. Missing measured states were inserted and inverse and mirror transitions were also taken into consideration. If there appeared a level in experimental compilations without definite spin and parity assignment, we did not replace (insert) theoretical levels with the experimental ones beyond this excitation energy. In our calculations, we summed the partial rates over 200 initial and as many final states (to ensure satisfactory convergence) to get the total capture rate. For details we refer to [19].

Realizing the pivotal role played by ⁵⁵Co and ⁵⁶Ni for the core collapse, Langanke and Martinez-Pinedo also calculated these electron capture rates separately [24]. They used the shell model diagonalization technique in the *pf* shell using the KB3 interaction [25] for their calculations. Due to model space restrictions and number of basis states involved in their problem, [24] performed the calculation only for the ground state of $^{56}$Ni. For $^{55}$Co two excited states (2.2 MeV and 2.6 MeV) along with the ground state were considered for calculations.

We did compare our B(GT) strength functions in the iron mass region with the experimental values and found satisfactory agreement. For details we refer to [19]. Normally in shell model calculations emphasis is laid more on interactions as compared to correlations. With QRPA, the story is other way round. In this Letter we compare the two different microscopic approaches.

The GT strength distributions for the ground state and two excited states in $^{55}$Co are shown in Fig. 1, whereas Fig. 2 shows a similar comparison for the ground state of $^{56}$Ni. Here we also compare our calculations with those of [24]. The upper panel shows our results as compared to the results of [24] (lower panel).

We note that our GT strength is fragmented over many daughter states. At higher excitation energies, E > 2.5 MeV, the calculated GT strengths represent centroids of strength (distributed over many states). We observe from our calculations that for the ground state of $^{55}$Co, the GT centroid resides in the energy range, E = 7.1 - 7.4 MeV in the daughter $^{55}$Fe, and it is, more or less, around E = 6.7 – 7.5 MeV for the excited states. There is one GT strength peak at 11.6 MeV in the ground state of $^{55}$Co, and similar peak for the GT strength is also observed in excited states around the same energy domain. For $^{56}$Ni, we calculate the total GT strength, from the ground state, to be 8.9 ([24] reported a value of 10.1 and Monte Carlo shell model calculations resulted in a value of 9.8 ± 0.4 [26]). Our corresponding value for the case of $^{55}$Co is 7.4 as compared to the value 8.7 reported by [24].

Our electron capture rates for $^{55}$Co and $^{56}$Ni are shown in Figs. 3 and 4, respectively. The temperature scale $T_9$ measures the temperature in $10^9$ K and the density shown in the legend has units of g cm$^{-3}$. We calculate these rates for densities in the range 10 to $10^{11}$ g/cm$^3$. Fig. 3 and Fig. 4 show results for a few selected density scales. These figures depict that for a given density, the electron capture rates remain, more or less, constant for a certain temperature range. Beyond a certain shoot off temperature the electron capture rates increase approximately linearly with increasing temperature. This rate of change is independent of the density (till $10^7$ g cm$^{-3}$). For higher density, $10^{11}$ g cm$^{-3}$ (density prior to collapse), we note that the linear behaviour starts around $T_9 = 10.0$. The region of constant electron capture rates, in these figures, with increasing temperature, shows that before core collapse the beta-decay competes with electron capture rate.

At later stages of the collapse, beta-decay becomes unimportant as an increased electron chemical potential, which grows like $\rho^{1/3}$ during infall, drastically reduces the phase space. This results in increased electron capture rates during the collapse making the matter composition more neutron-rich. Beta-decay is thus rather unimportant during the collapse phase due to the Pauli-blocking of the electron phase space in the final state.

How do our rates compare with those of [24]? The comparison is shown in Fig. 5 and Fig. 6 for $^{55}$Co and $^{56}$Ni, respectively. Here the right panel shows the rate of [24]. Our rates are depicted in the left panel. These calculations were performed for the same temperature and density scale as done by [24]. $\rho_7$ implies density in units of $10^7$ g cm$^{-3}$ and $T_9$ measures temperature in $10^9$ K.

For $^{55}$Co, our rates are much stronger and differ by almost two orders of magnitude at low temperatures as compared to those of [24]. At higher temperatures our rates are still a factor of two more than those of [24].

For the other interesting case, $^{56}$Ni, the story is different. Here at low temperatures our rates are still enhanced (by a factor of 4 at low temperatures and densities). At intermediate temperatures and density scales we are in good agreement and then at high temperatures and densities, shell model rates surpass our rates (by as much as a factor of 3). The difference decreases with increasing density. Collapse simulators should take note of our enhanced rate at presupernova temperatures. We took into consideration low-lying parent excited states in our rate calculations without assuming the so-called Brink's hypothesis (which states that the GT strength distribution on excited states is identical to that from the ground state, shifted only by the excitation energy of the state).

What implications do these rates have on the dynamics of core collapse? The nuclei which cause the largest change in $Y_e$ are the most abundant ones *and* the ones with the strongest rates. Incidentally, the most abundant nuclei tend to have small rates (they are more stable) and the most reactive nuclei tend to be present in minor quantities.

Our calculation certainly points to a much more enhanced capture rates as compared to those given in [24]. The electron capture rates reported here can have a significant astrophysical impact. According to the authors in [11], $\dot{\Psi}_e$ (rate of change of lepton-to-baryon ratio) changes by about 50% due to electron capture on $^{55}$Co (and about 25% for the case of $^{56}$Ni). It will be very interesting to see if these rates are in favor of a prompt collapse of the core. We also note that authors in [3] do point towards the fact that the spherically symmetric core collapse simulations, taking into consideration electron capture rates on heavy nuclides, still do not explode because of the reduced electron capture in the outer layers slowing the collapse and resulting in a shock radius of slightly larger magnitude. We are in a process of finding the affect of inclusion of our rates in stellar evolution codes and hope to soon report our results.

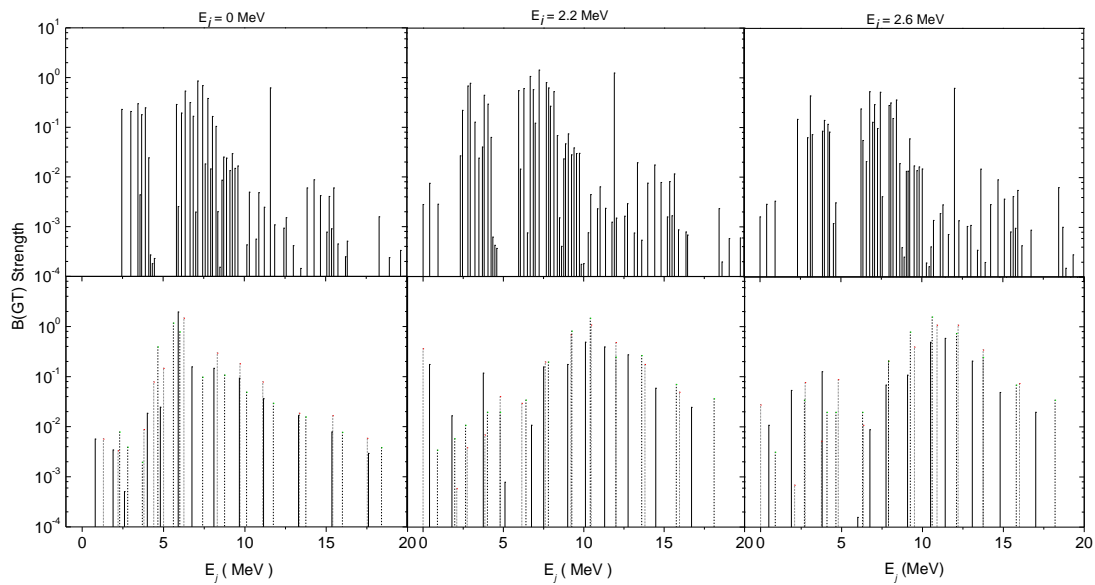

Fig. 1. Gamow-Teller (GT) strength distributions for $^{55}$Co. The upper panel shows our results of GT strength for the ground and first two excited states. The lower panel shows the results for the corresponding states calculated by [24]. $E_i$ ($E_j$) represents parent (daughter) states.

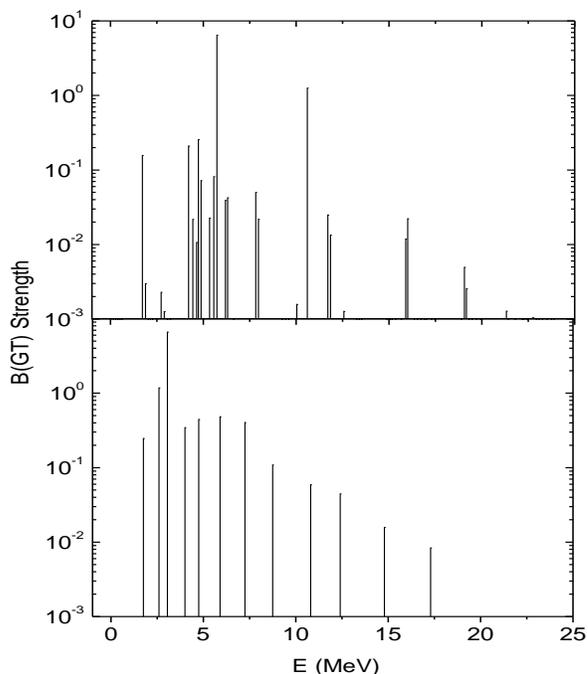

Fig. 2. Gamow-Teller distribution for $^{56}$Ni ground state. For comparison the calculated GT strength by [24] is shown in the lower panel. Here the energy scale refers to excitation energies in the daughter nucleus.

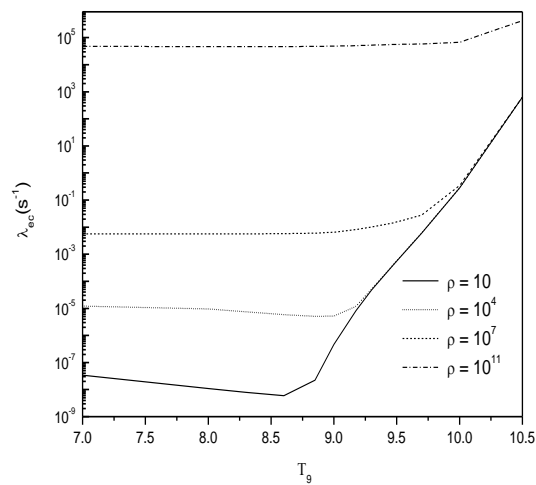

Fig. 3. Electron capture rates on $^{55}$Co as function of temperature for different selected densities. For units see text.

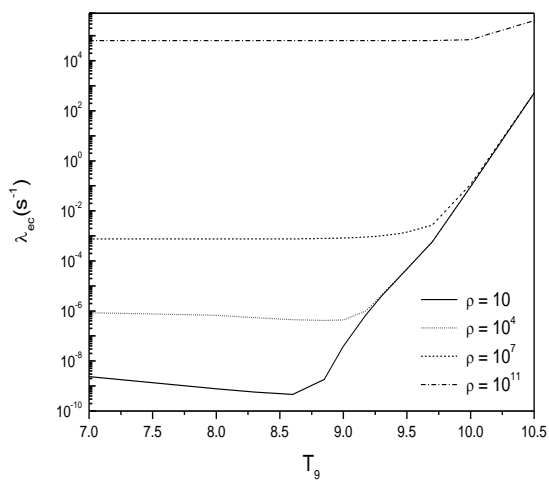

Fig. 4. Electron capture rates on $^{56}$Ni as function of temperature for different selected densities. For units see text.

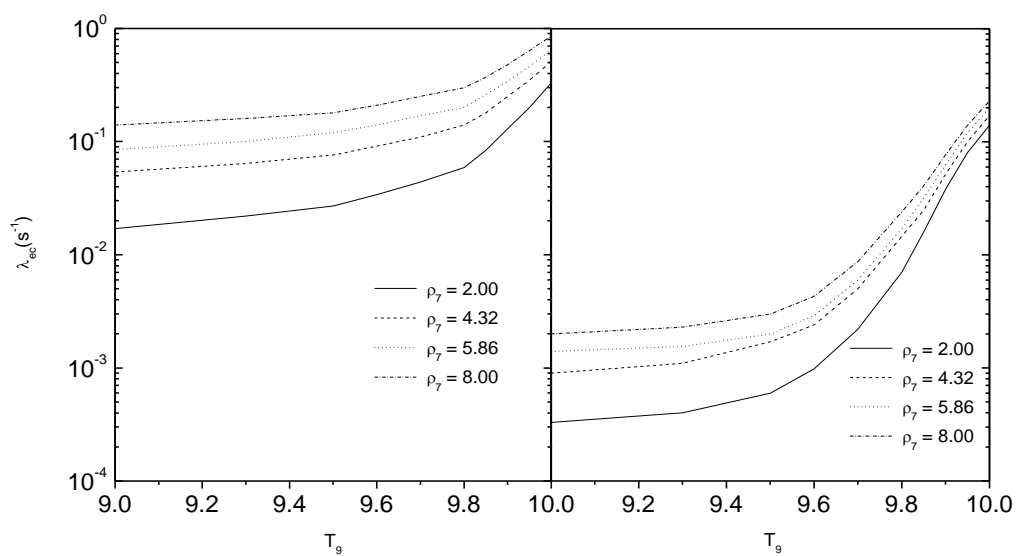

Fig. 5. Electron capture rates on $^{55}$Co as function of temperature for different densities (left panel). The right panel shows the results of [24] for the corresponding temperatures and densities. For units see text.

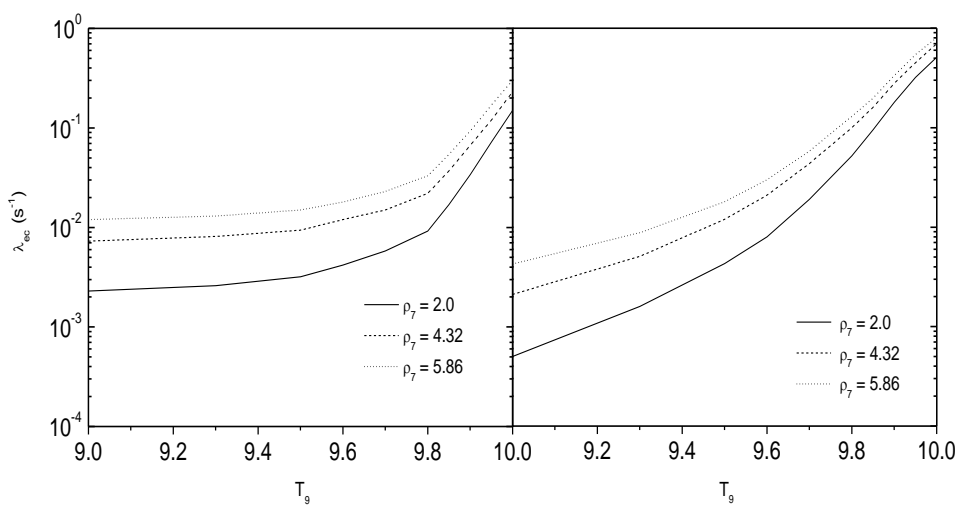

Fig. 6. Electron capture rates on $^{56}$Ni as function of temperature for selected densities (left panel). The right panel shows the results of [24] for comparison. For units see text.